\documentclass[journal]{IEEEtran}
\usepackage{cite}
\usepackage{amsmath,amssymb,amsfonts}
\usepackage{algorithmic}
\usepackage{graphicx}
\usepackage{textcomp}
\usepackage[ruled,linesnumbered]{algorithm2e}
\usepackage[affil-it]{authblk}
\usepackage{blindtext}
\usepackage{CJK}
\usepackage{type1cm}
\usepackage{times}
\usepackage[marginal]{footmisc}
\usepackage[affil-it]{authblk}
\usepackage{color}
\usepackage{multirow}
\usepackage{comment}
\usepackage{diagbox}
\usepackage{siunitx}
\usepackage{lipsum}
\usepackage{cuted}
\usepackage{amsmath}
\usepackage{multirow}
\usepackage{amssymb}
\usepackage{nomencl}
\usepackage{multirow}
\makenomenclature
\usepackage{etoolbox}
\renewcommand\nomgroup[1]{%
  \item[\bfseries
  \ifstrequal{#1}{A}{Abbreviations}{%
  \ifstrequal{#1}{B}{Variables}{%
  \ifstrequal{#1}{C}{Subscripts and Superscripts}{}}}%
]}

\begin{document}
\title{Sensor Attacks and Resilient Defense on HVAC Systems for Energy Market Signal Tracking}

\author{Guanyu~Tian,
        Qun~Zhou Sun,
        Yiyuan Qiao
        \thanks{
        \indent Guanyu Tian is with the Department of Marine Engineering Technology, Texas A\&M University at Galveston, Galveston, TX 77554 USA (email: tiang@tamug.edu)
        
        Qun Zhou Sun and Yiyuan Qiao are with the Department of Electrical and Computer Engineering, University of Central Florida, Orlando, FL 32816 USA (e-mail: QZ.Sun@ucf.edu; yiyuan.qiao@ucf.edu)
        }
        }

\date{\today}

\maketitle
\begin{abstract}
The power flexibility from smart buildings makes them suitable candidates for providing grid services. The building automation system (BAS) that employs model predictive control (MPC) for grid services relies heavily on sensor data gathered from IoT-based HVAC systems through communication networks. However, cyber-attacks that tamper sensor values can compromise the accuracy and flexibility of HVAC system power adjustment. Existing studies on grid-interactive buildings mainly focus on the efficiency and flexibility of buildings' participation in grid operations, while the security aspect is lacking. In this paper, we investigate the effects of cyber-attacks on HVAC systems in grid-interactive buildings, specifically their power-tracking performance. We design a stochastic optimization-based stealthy sensor attack and a corresponding defense strategy using a resilient control framework. The attack and its defense are tested in a physical model of a test building with a single-chiller HVAC system. Simulation results demonstrate that minor falsifications caused by a stealthy sensor attack can significantly alter the power profile, leading to large power tracking errors. However, the resilient control framework can reduce the power tracking error by over 70\% under such attacks without filtering out compromised data.
\end{abstract}


\IEEEpeerreviewmaketitle

\section{Introduction} \label{intro}
\IEEEPARstart{T}{he} involvement of building HVAC systems in power grid operations dates back to the 1970s, when demand response programs were first introduced in the US. Initially, the programs focused on interruptible electric service, which permitted utilities to interrupt the power supply temporarily during peak demand periods \cite{parmesano1983evolution}. In the subsequent decades, more advanced demand response programs emerged, enabling utilities to directly control the operation of HVAC systems in grid-interactive buildings \cite{eto1996past}. Customers that comply with demand response signals can receive financial incentives \cite{rahimi2010demand, su2009quantifying}. In 2011, FERC order 745 was issued to allow demand-side resources to participate in ancillary service markets. With the approval of FERC order 2222 in 2020, grid-interactive buildings are officially allowed to bid power reduction into the energy market through aggregators. To participate in the energy market, the load profile of a grid-interactive building must be flexibly and accurately controlled, so that it can track the dispatched power signals from the system operators and maintain market participation qualification \cite{wang2013electrical}. 

Many control strategies have been proposed for building HVAC systems to track power signals, including the feedback control schemes that control the supply air fan for providing frequency regulation services \cite{beil2016frequency, hao2014ancillary}, and the model predictive control (MPC) methods that can achieve comprehensive demand response services and participate in energy markets \cite{oldewurtel2012use, bianchini2016demand, tang2019model}. 


All controls need input from HVAC sensors, whose values are prone to cyber attacks. Ideally, MPC methods can yield perfect control performance for providing demand response services given all accurate inputs. However, the building automation system (BAS) that realizes the MPC is a centralized controller that relies heavily on sensor data collected from the IoT-based HVAC system through communication networks \cite{minoli2017iot}. Due to the uncertainty of measurement error and the vulnerability of IoT devices, the actual power tracking performance of MPC-based HVAC system control can potentially be compromised by sensor offset caused by faults or cyber-attacks \cite{yoon2019impacts}. 

With more and more buildings participating in grid operations, the vulnerability of building BAS opens the door to cyber attacks targeting power systems. Existing power system cyber attacks mainly focus on the system-wide impact of attacked aggregated power. For instance, the supervisory control and data acquisition (SCADA) attacks on transmission system state estimation have been studied in \cite{ten2008vulnerability, zhang2016inclusion}, where the falsified signals are the power consumption at PQ nodes. Similarly, for distribution systems, attacks on advanced metering infrastructure (AMI) also affect system operations by sending false power consumption signals \cite{jiang2015outage}. The load-altering attacks that can actually alter the load power have been identified in \cite{amini2016dynamic}, which can potentially yield more severe consequences, such as system instability. Although the system-wide impact and defense strategies have been extensively studied, it is necessary to defend against such attacks at their root. Unfortunately, the mechanism behind the load-altering attacks is not well understood, let alone the corresponding defense strategies.



In this paper, we study the physical model of HVAC systems, and investigate potential attacks by breaching the sensors. The designed cyber attack could potentially deteriorate the performance of grid-interactive buildings when tracking grid dispatch signals. It is a stealthy attack that is undetected by bypassing building fault detection rules. Then, a robust control method is proposed to withstand such an attack. First, the sensor attack is assumed to be launched by the worst-case probabilistic attack within the presumed ambiguity set that mimics the set of building fault detection and diagnosis rules. The attacking objective is to maximize the expected power tracking error while only introducing unnoticeable sensor value falsifications. Then, a resilient control method against such attacks is proposed. Considering the uncertainty of falsified sensor value under the stealthy sensor attack, the resilient control is initially formulated as a two-level distributionally robust optimization problem. Then the initial intractable formulation is relaxed to a single-level optimization problem that can be solved with high time efficiency.

The contributions of this paper are four folds:  
\begin{enumerate}
\item \textbf{We are among the first few to consider the impact of a cyber attack on HVAC controls on the grid operations.} The lack of smart building cyber security studies motives more in-depth research on how building cyber security affects its grid services. The research is critically needed because sensor falsification attacks can significantly deviate the power consumption of HVAC systems from the target power and compromise the quality of grid services.
\item \textbf{We designed a stealthy attack model that considers building dynamics while bypassing building fault detection rules.} The proposed attack model solves the worst-case distributions of attack signals that maximize the expected power tracking error. The feasible region for candidate distributions is an ambiguity set defined by a set of constraints that guarantees stealthiness.
\item \textbf{The robust control is proposed to accommodate the worst case of sensor value distortions so that the HVAC can still track grid signals in a wide range of undesired attack conditions in the buildings.} The min-max robust optimization-based robust control provides the upper bounds of power tracking error within the ambiguity set. The tractable formulation is obtained through strong duality.
\item \textbf{The final results are validated in a high-fidelity building model in Dymola.} The proposed resilient control reduces power tracking error by over 60\% under sensor attacks.
\end{enumerate}

This paper is organized as follows. Section \ref{formu} introduces the hierarchical control framework of grid-interactive buildings in energy markets. Section \ref{ptracking} introduces the standard formulations of HVAC power tracking and the corresponding sensor attack methods. Section \ref{defense} proposes the resilient control method against stealthy sensor attacks and the derived tractable form. Finally, both attack and defense methods are validated in section \ref{case} using the Dymola model of a single-building HVAC system under various test cases. A thermal fluid simulation model is also adopted to cross-validate the numerical simulation results.

\section{Problem Description}\label{formu}

\begin{figure}[h]
\begin{center}
{\includegraphics[width=0.4\textwidth]{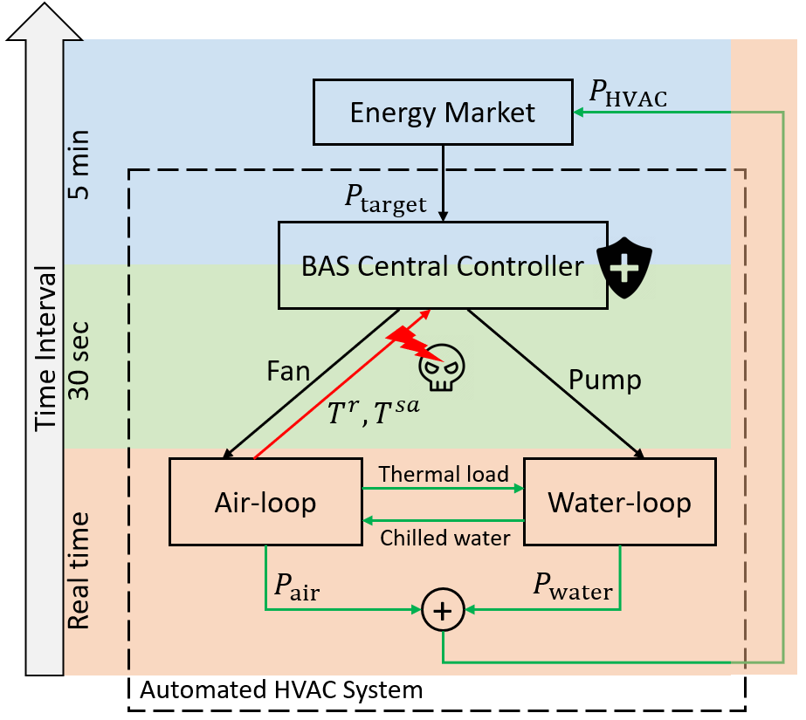}}
\end{center}
\caption{Grid-interactive building operations regarding time scale }
\label{overview}
\end{figure}

The flowchart in Fig. \ref{overview} demonstrates the operations of a grid-interactive building HVAC system participating in an energy market. The black arrows denote control signal flow, the red arrow denotes sensor data flow, and the green arrows denote the flow of physical variables in real time. Market signals are dispatched at 5-minute intervals. Upon receiving the target power dispatched from the market/system operators, the central controller of the building automation system (BAS) needs to determine control signals that can track the dispatched power profile while maintaining the room temperature within the comfort zone. The control signals are the air mass flow rate for fans and the water mass flow rate for pumps, which are updated every 30 seconds. The fan speed control signals are calculated upon receiving room temperature sensor values $T^r$ and supply air temperature sensor values $T^{sa}$. The water loop needs to provide enough cooling load for the air loop, hence, the control signal for pump speed is also dependent on the measurement of room temperature $T^r$ and supply air temperature $T^{sa}$. The overall HVAC power consumption mainly consists of fan power from the air loop, and pump and chiller power from the water loop.

Note that the sensor values are critical inputs to the BAS control for achieving desired power tracking performance. These sensors are usually unguarded IoT devices that are vulnerable to cyber-physical attacks. For instance, putting a heater or ice bag near the sensor can easily disturb the normal operations of an entire building \cite{tu2019trick}. In the cooling mode, if attackers maliciously increase the room temperature sensor value, the control signal of air mass flow rate sent to the fan will be higher than expected, yielding a higher fan speed and overall thermal load. Consequently, the overall HVAC power consumption will be higher. Similarly, if attackers change the supply air temperature sensor to a higher value, it can mislead the central controller to turn up the chill water mass flow rate to restore the supply air temperature back to the nominal value, leading to a higher pump and chiller power. Thus, falsifying sensor values can be an effective and low-cost attack method to alter HVAC power and compromise the power-tracking functionality of grid-interactive buildings. This can potentially disqualify buildings' participation in the energy market. 


Building sensor attacks can be launched in various ways through multiple channels, so it is particularly challenging to defend against all possible attacks. From attackers' perspective, the sensor falsification cannot exceed a reasonable region, otherwise, it can be easily detected by building fault data detection (FDD) algorithms. Also, these attack methods share the same objective of compromising HVAC power tracking performance. Exploiting these two facts, robust control can be leveraged in the control layer to defend against a variety of sensor attack methods. In this paper, we minimize the power tracking error under the worst-case attack within a statistically feasible region, that the upper bound of power tracking error is minimized. The robust control considers stealthy attacks within an ambiguity set bounded by FDD rules. The control is then reformulated in a tractable moment-based formulation solved through dualization. The robust formulation is then demonstrated in case studies to be very effective in protecting the market performance of buildings against sensor attacks.

\section{Sensor Attacks against HVAC Power Tracking}\label{ptracking}
\subsection{Power Tracking in Energy Markets}
Being able to accurately track power profiles dispatched by power system operators is a critical requirement for market participants. However, commercial buildings equipped with conventional HVAC systems with feedback control have very limited capacity to simultaneously achieve power tracking and temperature maintenance \cite{morocsan2010building}. Thus, model predictive control (MPC) is widely adopted in the existing literature for HVAC system power tracking \cite{oldewurtel2012use, bianchini2016demand, tang2019model, wang2022control}. 

The formulation in \eqref{DOformu} shows a standard MPC of an HVAC power tracking problem. 
\begin{itemize}
    \item {Standard Power Tracking Formulation:}
\end{itemize}
\begin{subequations}
\allowdisplaybreaks
\begin{align}
\min_{\dot{m}^{sa}} &\sum_{t=1}^T\left(P_t - P^{ref}_t\right)^2\\
s.t.\quad& P_t = b_1\dot{m}^{sa}_t + b_2\dot{m}^{sa}_tT^r_t + b_3\dot{m}^{sa}_tT^{sa}_t, \forall t\label{pmodel}\\
&T^r_{t+1}= c_0 + c_1T^r_t + c_2 \dot{m}^{sa}_{t}T^r_t + c_3\dot{m}^{sa}_{t}T^{sa}_t, \forall t\label{Tmodel}\\
&T^r_{lb}\leq T^r_t\leq T^r_{ub}, \forall t \label{tbounds}
\end{align}
\label{DOformu}
\end{subequations}
The objective is to minimize the squared error between the actual power profile $P$ and the market-dispatched power profile $P^{ref}$ within the look-ahead time window $T$. The decision variable is air mass flow rate $\dot{m}^{sa}_t$ at each time step $t$, which controls the speed of the supply air fan. The HVAC power $P$ and zone temperature $T^r$ are described by two physics-based equations, represented by \eqref{pmodel} and \eqref{Tmodel} respectively, with time-varying parameters. Detailed derivations of the two equations are given in Appendix A, and briefly discussed below. 

\begin{itemize}
    \item {HVAC Power Model:} \eqref{pmodel} is a linear function of the decision variable $\dot{m}^{sa}_t$, when the sensor inputs $T^r_t$ and $T^{sa}_t$ are known. The power is modeled based on HVAC system and environmental parameters, including the energy efficiency of chiller denoted by Coefficient of Performance (COP), outdoor air temperature $T_t^o$, and damper position $\beta$ that affects the ratio of outdoor air into the building. $c_{air}$ denotes the specific heat capacity of air.  
    \item {Thermal Zone Temperature Model:} \eqref{Tmodel} is discretized from the nodal current balance of the RC equivalent model, which considers the building parameters of thermal resistance $R$ and thermal capacitance $C$. $\Delta t$ denotes the time interval of discretization. $c_0$ represents a constant term that is independent of the control variable $\dot{m}^{sa}_t$ and the vulnerable sensor measurement $T^r_t$ and $T^{sa}_t$. 
\end{itemize}

For simplicity, the constant parameters are compressed into coefficients $b_1-b_3$ and $c_0-c_3$ (see Appendix A). \eqref{tbounds} is the security constraint of HVAC operation that maintains the predictive zone temperature profile within the bounds of comfort zone $\left[T^{lb}, T^{ub}\right]$.

\subsection{Sensor Attacks}
Grid-interactive smart buildings, regarded as a typical cyder-physical system, can be attacked from both physical and cyber layers. For example, attacks from the physical level can be easily launched. Unlike power control centers, most buildings do not have a high security level, and rooms are often unguarded. The sensors of room temperature $T^r$ and supply air temperature $T^{sa}$ are placed inside the rooms and air vents, which can be easily accessed. Falsifying these sensor values is relatively simple, such as by covering the sensors with insulation materials. The impact of the simple physical attack can be formulated as \eqref{naive}, where the falsified sensor value under attack is shifted by an offset $e_a$ from the actual temperature $T_{a}$. Imposing the shifted values $T_a^r$ and $T_a^{sa}$ into \eqref{DOformu} can lead to inaccurate modeling and thus deviates the actual HVAC power from the target profiles.
\begin{equation}
\allowdisplaybreaks
\begin{aligned}
T_a^r &= T^r+e_a^r\\
T_a^{sa} &= T^{sa}+e_a^{sa}
\end{aligned}
\label{naive}
\end{equation}

A cyber-layer attack can also be launched. Given the distributed and long-distance placement of sensors in a commercial building, it becomes necessary to transmit sensor measurements to the central controller of the Building Automation System (BAS). This transmission is achieved through the utilization of a local area network (LAN), which typically implements standardized communication protocols such as BACnet, KNX, Modbus, and others \cite{cash2023false}. The security of communication links is a critical concern, as the protocols are not usually designed with security considerations. Despite efforts to enhance cyber security through measures like firewalls and encryption, it has been observed that even BAS systems in highly advanced companies like Google \cite{google} can still be compromised. This implies communication links are vulnerable and can be breached during transmission between sensors and central controllers. 

Attackers may also intelligently falsify specific sensor values to achieve their malicious goals. For example, \eqref{rmse0} demonstrates an attacking objective function, where the attacking goal is to maximize the power tracking error within the look-ahead window by falsifying the sensor values in a coordinated manner. 
\begin{equation}
\allowdisplaybreaks
\max_{T^r,T^{sa}}\sum_{t=1}^N \left(P^{HVAC}_t-P^{ref}_{t}\right)^2
\label{rmse0}
\end{equation}

The decision variables are the attack signals for the room temperature and supply air temperature sensors at every time step. Thus, the optimal solution is the combination of them that coordinates the spatial relationship between the two sensors and the temporal relationships between the time steps. The power tracking error is evaluated using the sum of the second-order error between the actual power and the reference power at every step, such that the root-mean-square error (RMSE), a common metric for evaluating trajectory distance, can be maximized, indicating the most malicious attack is achieved.



\subsection{Bypassing Fault Detection Rules}

Modern buildings are equipped with fault data detection (FDD) algorithms, especially for their core components, such as air-handling units (AHUs). To ensure its safe and reliable operations and to identify faulty parts under abnormal conditions, sensors are deployed at critical nodes for monitoring and fault detection purposes. The room temperature is measured from the thermostat located within the room, the sensor near the vent measures the supply air temperature, and the sensor inside the mixing box measures the mixed air temperature. Moreover, the outdoor air temperature is measured from a temperature sensor placed outside the building. Abnormal sensor data inputs may trigger FDD rules and issue alarms to building managers. Thus, bypassing these FDD criteria is necessary for launching stealthy sensor attacks.

The air-handling-unit performance assessment rules (APAR) is a set of fault identification rules that has been widely adopted in existing commercial HVAC systems \cite{house2001expert}. APAR is generalizable to buildings of different sizes and operating modes. These rules are intentionally designed with a certain level of error tolerance to ensure detection reliability. However, this tolerance inadvertently creates an opportunity for subtle sensor falsification attacks to bypass detection.

APAR consists of a total of 28 rules that describe all possible abnormal conditions across 8 different operation modes. Among these rules, 20 are specifically designed to identify sensor errors. Among the operation modes, mode 3 and mode 4 are associated with mechanical cooling modes, which use chilled water to cool the room down when the outside air temperature is higher than the room temperature. As the power flexibility of the HVAC system primarily stems from the chiller, which is exclusively utilized for mechanical cooling, the focus of this paper lies on the overlapped rules: \#8, \#10, \#11, \#12, \#16, \#17, and \#18, which pertain to sensor errors under mechanical cooling modes. The satisfaction of any rules below would trigger fault detection alarm. 
\begin{itemize}
    \item {Rule 8:} $T^{o}< T^{sa} - \Delta T^{sf}-\varepsilon_t$
    \item {Rule 10:} $\mid T^{o} - T^{mix} \mid > \varepsilon_t$
    \item {Rule 11 \& 16:} $T^{sa}> T^{mix}+\Delta T^{sf} + \varepsilon_t$
    \item {Rule 12 \& 17:} $T^{sa}> T^{r}-\Delta T^{rf} + \varepsilon_t$
    \item {Rule 18:} $\mid T^{r} - T^{o}\mid \geq \Delta T_{min}$
\end{itemize}
where $\Delta T^{sf}$ and $\Delta T^{rf}$ denote the temperature change across the supply fan and return fan due to the fan motor heat gain. $T^{mix}$ denotes the mixed air temperature formulated in \eqref{mixbox}, which is the weighted average of return air $T^{ra}$ and the outside air $T^o$. The weight $\beta$ is the opening ratio of the damper. $\varepsilon_t$ is a small error tolerance term, and $\Delta T_{min}$ is the threshold temperature gap that prevents the ratio of outside air entering the AHU from being too high or too low.

Under mechanical cooling modes, the cooling coil that contains the chill water is in use. The temperature of the mixed air entering the cooling coil must be higher than the supply air left it. Thus, in normal conditions, the supply air temperature should be lower than the mixed air temperature. Considering the temperature gain from the supply air fan $\Delta T^{sf}$ is positive, we have $T^{sa}<T^{mix}<T^{mix}+\Delta T^{sf} + \varepsilon_t$, and rule 11 \& 16 are only true when any temperature sensors are faulty. Moreover, the supply air temperature must be lower than the room temperature to be able to cool it down, therefore, rule 12 \& 17 also reflect sensor value inconsistency. Rule 18 is an abnormal ventilation condition that reflects the fraction of outside air entering the thermal zone is either too high or too low.

The mechanical cooling with 100\% outside air mode is a special case, where the mixing box damper is fully open $\left(\beta=100\%\right)$. Hence, the mixed air is all from the outside, indicating the mixed air temperature should be approximately the same as the outside air. Thus, when their difference is larger than a certain threshold, it is considered to be a sensor inconsistency condition by rule 10. Given this fact, the supply air, which is supposed to be lower than the mixed air, should also be lower than the outside air as well. This type of sensor inconsistency is reflected by rule 8. 

\begin{figure}[h]
\begin{center}
{\includegraphics[width=0.45\textwidth]{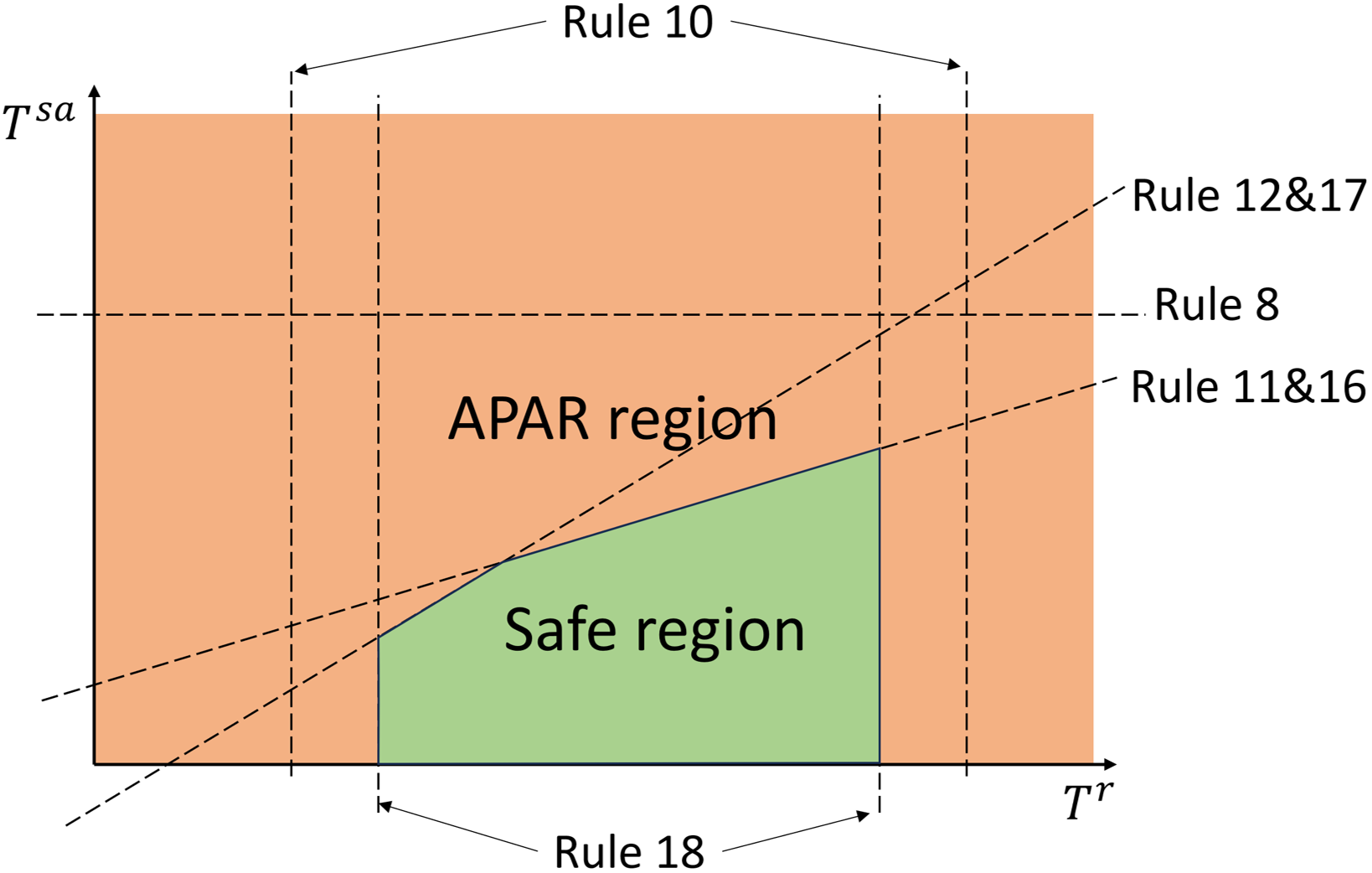}}
\end{center}
\caption{Faulty and safe areas under APAR detection}
\label{aparregions}
\end{figure}

Fig. \ref{aparregions} demonstrates that the above rules construct a linear polyhedron, noted as the APAR region, with respect to the room temperature $T^r$ and supply air temperature $T^{sa}$ that is regarded as abnormal conditions. Correspondingly, the complement region is the safe region for attackers to launch stealthy sensor attacks, as formulated in \eqref{saferegion}.  

\begin{equation}
\resizebox{.9\hsize}{!}{$
\left[\begin{aligned}
&\beta-1     & 1 \\
&-1 & 1 \\
&1       & 0\\
&-1       & 0\\
&0       & 1\\
&0       & -1
\end{aligned}\right] \left[\begin{aligned}
&T^r \\
&T^{sa}
\end{aligned}\right] \leq \left[\begin{aligned}
&\beta T^o+\Delta T_{sf} + \varepsilon_t \\
&-\Delta T_{rf} + \varepsilon\\
& T^o + \Delta T_{min}\\
& -T^o - \Delta T_{min}\\
& \min\left(T^o+\Delta T_{sf}+\varepsilon_t,T^o+\frac{\varepsilon_t}{1-\beta}\right)\\
&-T^o+\frac{\varepsilon}{1-\beta}
\end{aligned}\right]
$}
\label{saferegion}
\end{equation}

As long as the falsified sensor values stay within this safe region, APAR cannot detect them. This condition can be easily satisfied by adopting the safe region \eqref{saferegion} as a constraint in the attack model \eqref{rmse0}. Such attacks are referred to as stealthy attacks. 


\section{Resilient Control Against Worst-case Stealthy Sensor Attacks}\label{defense}

Stealthy sensor attacks introduce errors in the estimation of thermal load, which can be accumulated over time and affect HVAC power. Thus, a resilient control that can robustly track the desired load profile against stealthy sensor attacks is critical to HVAC control for providing high-quality grid services and energy market participation. 


\subsection{Resilient Control with Practical Attacking Considerations}
From the defenders' perspective, the settings of the safe region are known, while the detailed attack formulations are unknown and can be versatile. It is unlikely to enumerate all possible attack methods and design defense strategies accordingly. Thus, a robust power tracking control method that is effective under many possible attacks is desired. This requirement is in accordance with the feature of robust optimization, which finds the control decision that achieves the minimum error under the worst-case scenario. The solution obtained from the two-level min-max robust optimization provides the upper bound of error under possible attack scenarios. However, in our case, the worst-case scenarios are always obtained at the vertex of the safe region, which is too conservative to be practical.

Given the fact that stealthy attack signals are expected to fall inside the safe region in a natural manner, we propose the resilient control formulation with practical attacking considerations in \eqref{minmax0}.
\begin{subequations}
\allowdisplaybreaks
\begin{align}
\min_{\dot{m}^{sa}}\quad&\max_{T^r\sim\mathbb{P},T^{sa}\sim\mathbb{Q}}\mathbb{E}_{\mathbb{P},\mathbb{Q}}\left[\sum_{t=1}^N \left(P^{HVAC}_t-P^{ref}_{t}\right)^2\right]\label{rmse2}\\
& s.t \quad \eqref{pmodel}-\eqref{tbounds} \notag\\
& \quad \quad \left(\mathbb{P},\mathbb{Q}\right)\in\mathcal{S}\\
s.t \quad & \dot{m}^{sa}_{lb}\leq\dot{m}^{sa}_t\leq\dot{m}^{sa}_{ub}, \forall t
\end{align}
\label{minmax0}
\end{subequations}

The lower-level maximization problem formulates a probabilistic worst-case scenario, where the attack signals $T^r$ and $T^{sa}$ are sampled from normal distributions $\mathbb{P}$ and $\mathbb{Q}$ respectively. The stealthiness is maintained by limiting the candidate distributions within the ambiguity set $\mathcal{S}$ formulated in \eqref{ambiguityset}, such that the sampled attack signals fall inside the safe region with a high probability $1-\alpha$.  
\begin{equation}
\allowdisplaybreaks
\begin{aligned}
\mathcal{S}=\{\left(\mathbb{P},\mathbb{Q}\right)|&Pr \left(\hat{T}^r_{\mathbb{P}}\in[T^r_{ref}-\varepsilon,T^r_{ref}+\varepsilon]\right)\geq 1-\alpha,\\
&VAR\left[\mathbb{P}\right]\leq\sigma^2_{T^r},\\
&Pr \left(\hat{T}^{sa}_{\mathbb{Q}}\in[T^{sa}_{ref}-\gamma,T^{sa}_{ref}+\gamma]\right)\geq 1-\alpha,\\
&VAR\left[\mathbb{Q}\right]\leq\sigma^2_{T^{sa}}\}
\end{aligned}
\label{ambiguityset}
\end{equation}

It is worth mentioning that, the deterministic attack scenarios are also covered by the probabilistic scenarios as special cases, where the variance is zero and the sampled attack signal is always the mean value. In this way, the proposed probabilistic worst-case scenario-based resilient control is robust against both deterministic and probabilistic stealthy sensor attacks. 

\subsection{Moment-based Formulation} 
The resilient control \eqref{minmax0} is not a tractable formulation that can be solved efficiently by off-the-shelf solvers. To reduce problem complexity and enhance the applicability of the proposed defense strategy, the lower-level stochastic optimization problem is reformulated into a moment-based deterministic convex optimization problem. The decision variables are substituted by $\left[\mathbb{E}\left[T^{r}_{t}\right] \mathbb{E}[T^{r^2}_{t}] \mathbb{E} \left[T^{sa}_{t} \right] \mathbb{E}[T^{sa^2}_{t}] \mathbb{E}\left[T^{r}_{t}T^{sa}_{t}\right]\right]^T$, where $\mathbb{E}\left[T^r_t\right]$ and $\mathbb{E}[T^{r^2}_t]$ denote the first and second moments of the falsified room temperature sensor attack distribution, and $\mathbb{E}\left[T^{sa}_t\right]$ and $\mathbb{E}[T^{sa^2}_t]$ denote the first and second moments of the falsified supply air temperature sensor attack distribution. $\mathbb{E}\left[T^{r}_tT^{sa}_t\right]$ denotes the expected value of the co-variant term that considers the joint distribution of the two sensors. 

The moment-based ambiguity set $\mathcal{S}$ and constraint \eqref{tbounds} are reformulated in \eqref{ambiguity} regarding the moments at the first time step. The remaining time steps are accounted for in \eqref{evolve1}. \eqref{tempconst} ensures that the projected room temperature is kept within the comfort zone under the worst-case, such that no occupant complaint issued or work orders regarding abnormal room temperature will be caused to compromise the stealthiness of attacks. Constraints \eqref{meanlimit1} and \eqref{meanlimit2} keep the mean of the worst-case distribution within the neighborhood of the true value. $\varepsilon$ and $\gamma$ are the pre-selected error tolerance. Constraints \eqref{varlimit1} and \eqref{varlimit2} limit the variance of the worst-case distribution from being too large, such that the sampled falsified sensor values can be less oscillatory. The left-hand side of both constraints denotes the upper bounds of variance. $\sigma_{T^r}$ and $\sigma_{T^{sa}}$ are the pre-selected tolerance on standard deviations. 
\begin{subequations}
\allowdisplaybreaks
\begin{align}
&T^r_{lb}\leq \mathbb{E}\left[T^{r}_{t}\right]\leq T^r_{ub}, \forall t \label{tempconst}\\
&\hat{T}^r_1-\varepsilon\leq \mathbb{E}\left[T^{r}_{1}\right]\leq \hat{T}^r_1+\varepsilon \label{meanlimit1}\\
&\hat{T}^{sa}_1-\gamma\leq \mathbb{E}\left[T^{sa}_{1}\right]\leq \hat{T}^{sa}_1+\gamma \label{meanlimit2}\\
&\mathbb{E}\left[T^{r^2}_{1}\right] - \left(\hat{T}^r_1-\varepsilon\right)^2\leq \sigma_{T^r}^2 \label{varlimit1}\\
&\mathbb{E}\left[T^{sa^2}_{1}\right] - \left(\hat{T}^{sa}_1-\gamma\right)^2\leq \sigma_{T^{sa}}^2 \label{varlimit2}
\end{align}
\label{ambiguity}
\end{subequations}

In order to address the unrealistic situations that arise from significant step changes, it is necessary to acknowledge the interdependence of the decision variables at different time steps. Consequently, it becomes crucial to consider their temporal relationship. The dynamics of the moments of room temperature distribution are captured in \eqref{evolve1}, which incorporates the zone temperature model \eqref{Tmodel}. This formulation allows for a comprehensive understanding of how the decision variables evolve over time, accounting for the influence of the zone temperature.
\begin{subequations}
\allowdisplaybreaks
\begin{align}
\mathbb{E}\left[T^{r}_{t+1}\right] =& \left(c_1+c_2m^{sa}_t\right)\mathbb{E}\left[T^{r}_t\right] + c_3m^{sa}_t\mathbb{E}\left[T^{sa}_t\right] + c_0\\
\mathbb{E}\left[T^{r^2}_{t+1}\right] =& \left(c_1+c_2m^{sa}_t\right)^2\mathbb{E}\left[T^{r^2}_t\right] + c_3^2m^{sa^2}_t\mathbb{E}\left[T^{sa^2}_t\right] \notag\\
&+ 2c_0\left(c_1+c_2m^{sa}_t\right)\mathbb{E}\left[T^{r}_t\right] + 2c_0c_3m^{sa}_t\mathbb{E}\left[T^{sa}_t\right] \notag\\
&+ 2\left(c_1+c_2m^{sa}_t\right)c_3m^{sa}_t\mathbb{E}\left[T^{r}_tT^{sa}_t\right] + c_0^2
\end{align}
\label{evolve1}
\end{subequations}

According to the cooling coil dynamic model, the supply air temperature mainly depends on the supply water temperature, which has a much longer time constant compared to the time interval and look-ahead time window of optimization formulation MPC. Hence, it is feasible to assume the first and second moments of supply air temperature distribution are static as formulated in \eqref{evolve2}.
\begin{subequations}
\allowdisplaybreaks
\begin{align}
\mathbb{E}\left[T^{sa}_{t+1}\right] =&\mathbb{E}\left[T^{sa}_{t}\right]\\
\mathbb{E}\left[T^{sa^2}_{t+1}\right] =&\mathbb{E}\left[T^{sa^2}_{t}\right]
\end{align}
\label{evolve2}
\end{subequations}

The dynamic model of the co-variant decision variable $\mathbb{E}\left[T^{r}T^{sa}\right]$ is formulated in \eqref{evolve3}, which is obtained from the combination of \eqref{evolve1} and \eqref{evolve2}.
\begin{subequations}
\allowdisplaybreaks
\begin{align}
\mathbb{E}\left[T^{r}_{t+1}T^{sa}_{t+1}\right] =& \left(c_1+c_2m^{sa}_t\right)\mathbb{E}\left[T^{r}_{t}T^{sa}_{t}\right] + c_3m^{sa}_t\mathbb{E}\left[T^{sa^2}_{t}\right] \notag\\
&+ c_0\mathbb{E}\left[T^{sa}_{t}\right]
\end{align}
\label{evolve3}
\end{subequations}

By substituting the power model \eqref{pmodel} into the objective function \eqref{rmse2}, the moment-based objective function is obtained in \eqref{attobj1}. The time series air mass flow rate $\dot{m}^{sa}$ is the upper-level problem decision variable, therefore the terms  $b_1^2\dot{m}^{sa^2}-2b_1P^{ref}_t\dot{m}^{sa}_t + P^{ref^2}_t$ are independent with the lower-level decision variables, i.e., the moments of attack distributions, and therefore are regarded as constants here.
\begin{subequations}
\allowdisplaybreaks
\begin{align}
\max_{}\sum_{t=1}^N & b_2^2\dot{m}^{sa^2}_t\mathbb{E}\left[T^{r^2}_t\right] + b_3^2\dot{m}^{sa^2}_t \mathbb{E}\left[T^{sa^2}_t\right] \notag\\
+&\left(2b_1b_2\dot{m}^{sa^2}_t-2b_2P^{ref}_t\dot{m}^{sa}_t\right)\mathbb{E}\left[T^{r}_t\right]\notag\\
+&\left(2b_1b_3\dot{m}^{sa^2}_t-2b_3P^{ref}_t\dot{m}^{sa}_t\right)\mathbb{E}\left[T^{sa}_t\right]\notag\\
+&2b_2b_3\dot{m}^{sa^2}_t\mathbb{E}\left[T^{r}_tT^{sa}_t\right]\notag\\
+&b_1^2\dot{m}^{sa^2}-2b_1P^{ref}_t\dot{m}^{sa}_t + P^{ref^2}_t \label{attobj1}\\
s.t. \qquad& \eqref{ambiguity}, \eqref{evolve1}, \eqref{evolve2}, \eqref{evolve3}
\end{align}
\label{momentformu}
\end{subequations}

Together with the moment-based constraints \eqref{ambiguity}-\eqref{evolve3}, the lower-level problem (i.e., attacker's problem) is now transformed into a deterministic convex optimization problem \eqref{momentformu}. The independent decision variables are the first and second moments at the first step, while the projected power tracking error is associated with the dependent moments at every time step that are related through the moment-based dynamic models.

\subsection{Solution Methodology through Dualization}
Considering that the moment-based formulation of the lower-level problem is convex, and by the strong duality theorem, its dual problem achieves the same optimal solution. Hence, the next step is to find the dual minimization problem and merge it with the upper-level problem to obtain the tractable formulation of the entire resilient control. 

For simplicity, we need to rearrange the primal problem into a compact formulation first. Let $x_t=\left[\mathbb{E}\left[T^{r}_{t}\right] \mathbb{E}[T^{r^2}_{t}] \mathbb{E}\left[T^{sa}_{t}\right] \mathbb{E}[T^{sa^2}_{t}] \mathbb{E}\left[T^{r}_{t}T^{sa}_{t}\right]\right]^T$, the equivalent matrix form of the lower-level problem can be obtained in \eqref{formu2}, where $c_t\in R^{5\times1}$, $A\in R^{5\times5}$, $B\in R^{5\times1}$, $C\in R^{8\times5}$, and $D\in R^{8\times1}$ are vectors and matrices constructed from previously introduced parameters and are summarized in Appendix B. 
\begin{subequations}
\allowdisplaybreaks
\begin{align}
\max_{x}&\sum_{t=1}^T c_tx_t+ \left(b_1^2\dot{m}^{sa^2}_t-2b_1P^{ref}_t\dot{m}^{sa}_t + P^{ref^2}_t\right)\\
s.t \quad & x_{t+1} = Ax_t + B, \forall t\\
&Cx_t+D\leq 0, \forall t
\end{align}
\label{formu2}
\end{subequations}

Furthermore, by introducing the slack variables $s$ for the inequality constraints and organizing the decision variables into $x' = \left[x_1^T x_2^T ... x_T^T s_1^T s_2^T ... s_T^T\right]^T$, the previous problem \eqref{formu2} can be equivalently reformulated into the compact formulation in \eqref{form2compact}, where the construction of $c'\in R^{7T+4\times}$, $A'\in R^{7T-1\times7T+4}$, and $B'\in R^{7T-1\times1}$ are summarized in Appendix B.
\begin{subequations}
\allowdisplaybreaks
\begin{align}
\max_{x'}\quad&c'^Tx'+\sum_{t=1}^T\left(b_1^2\dot{m}^{sa^2}_t-2b_1P^{ref}_t\dot{m}^{sa}_t + P^{ref^2}_t\right)\\
s.t \quad & A'x'+B' = 0
\end{align}
\label{form2compact}
\end{subequations}

Now, based on the compact formulation \eqref{form2compact}, we can construct the Lagrangian function \eqref{lagfunc}, where $\lambda$ denotes the vector of dual variables. 
\begin{equation}
\allowdisplaybreaks
\begin{aligned}
\mathcal{L}\left(x',\lambda\right) =& c'^Tx' + \lambda\left(A'x'+B'\right)\\
&+\left(b_1^2\dot{m}^{sa^2}-2b_1P^{ref}_t\dot{m}^{sa}_t + P^{ref^2}_t\right)\\
=& B'^T\lambda + \left(A'^T\lambda+c'\right)^Tx'\\
&+\sum_{t=1}^T\left(b_1^2\dot{m}^{sa^2}_t-2b_1P^{ref}_t\dot{m}^{sa}_t + P^{ref^2}_t\right)
\end{aligned}
\label{lagfunc}
\end{equation}

Then, the optimal solution of the original lower-level maximization problem can be obtained by finding the infimum of its dual problem \eqref{dual}, which can be merged with the upper-level minimization problem without loss of generality.
\begin{subequations}
\allowdisplaybreaks
\begin{align}
\min_{\lambda}\quad&B'^T\lambda+\sum_{t=1}^T\left(b_1^2\dot{m}^{sa^2}_t-2b_1P^{ref}_t\dot{m}^{sa}_t + P^{ref^2}_t\right)\\
s.t \quad & A'^T\lambda+c' \geq 0
\end{align}
\label{dual}
\end{subequations}

Considering \eqref{formu2} is a convex problem and strong duality holds, the proposed resilient control framework now is equivalently reformulated from the initial min-max robust optimization into the tractable single-level minimization problem \eqref{form3}, 
\begin{itemize}
    \item {Tractable Resilient Control Formulation:}
\end{itemize}
\begin{subequations}
\allowdisplaybreaks
\begin{align}
\min_{\dot{m}^{sa},\lambda} \quad& B'^T\lambda + \sum_{t=1}^T\left(b_1^2\dot{m}^{sa^2}_t-2b_1P^{ref}_t\dot{m}^{sa}_t + P^{ref^2}_t\right)\\
s.t \quad & A'^T\lambda+c' \geq 0\\
&\dot{m}^{sa}_{lb}\leq\dot{m}^{sa}_t\leq\dot{m}^{sa}_{ub}, \forall t
\end{align}
\label{form3}
\end{subequations}
where robust decisions can be made on supply air mass flow rate to prevent any adverse impact from attacks on HVAC power tracking energy market signals.

\section{Case Study}\label{case}
The proposed attack model and resilient control are tested using the digital twin models of a test building with a single-chiller HVAC system. The time span of the simulation is one day, and the simulations are carried out on a desktop with a 4-core 3.2 GHz CPU and 8 GB RAM. The target power profile is designed according to the baseline scenario power consumption and a $\pm15\%$ flexibility region. The baseline scenario assumes the HVAC is under temperature error feedback control. The standard MPC serves as the benchmark method, which works perfectly under accurate sensor values but yields large overall power profile deviations under sensor attacks. The power tracking error is effectively reduced by substituting the standard MPC with the proposed resilient MPC. It not only reduces the power tracking error by over 70\%, but also shows higher robustness against different levels of attack magnitudes. The computational requirement of solving the resilient control is sufficiently efficient for real-time applications. 

\subsection{Settings}
The performance of both the worst-case sensor attack and resilient control are tested under the simulated scenario where the daily operation of a single-building HVAC system is considered. The control decision making and the numerically predicted performances are conducted using Matlab scripts. The realistic performances are further validated using the physical model-based simulations in Dymola. The building and HVAC system parameters are summarized in Table \ref{para1}. 

\begin{table}[ht!]
\color{black}
\begin{center}
\caption{Thermal Zone \& HVAC Parameters}
\label{para1}
\scalebox{0.9}{
\begin{tabular}{c | c c | c c }
\hline
&Parameter & Value & Parameter & Value \\
\hline
\multirow{2}{*}{\shortstack{HVAC}}&$\beta$ &0.3&COP& 4.17\\
                                  &$T^{sa}_{nominal}$&\SI{24}{\celsius}&$T^{sa}_{nominal}$&\SI{16}{\celsius}\\
\hline
Thermal zone&$R$ &1e-5 \SI{}{\celsius}/W& $C$ & 3.6219e3J/\SI{}{\celsius}\\
\hline
Constants&$c_{air}$ &1014.54 J/(kg\SI{}{\celsius}) &$\Delta t$ &30 sec \\
\hline
\end{tabular}}
\end{center}
\end{table}

\begin{figure}[h]
\begin{center}
{\includegraphics[width=0.5\textwidth]{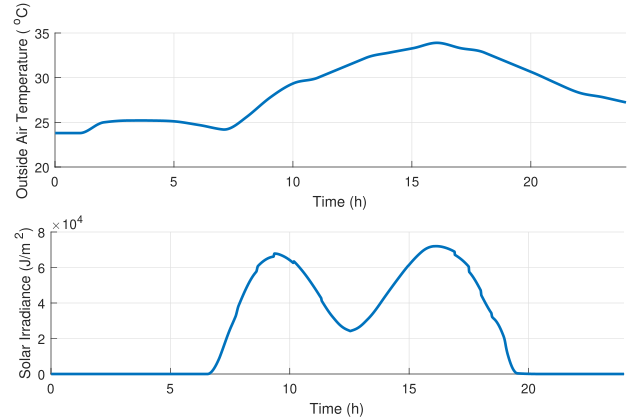}}
\end{center}
\caption{Environmental input profiles of the daily test case}
\label{env}
\end{figure}

The typical ambient temperature and solar irradiance profiles in the Orlando area, depicted in Fig. \ref{env}, are adopted as the environmental inputs of the test building system. 

\begin{figure}[h]
\begin{center}
{\includegraphics[width=0.5\textwidth]{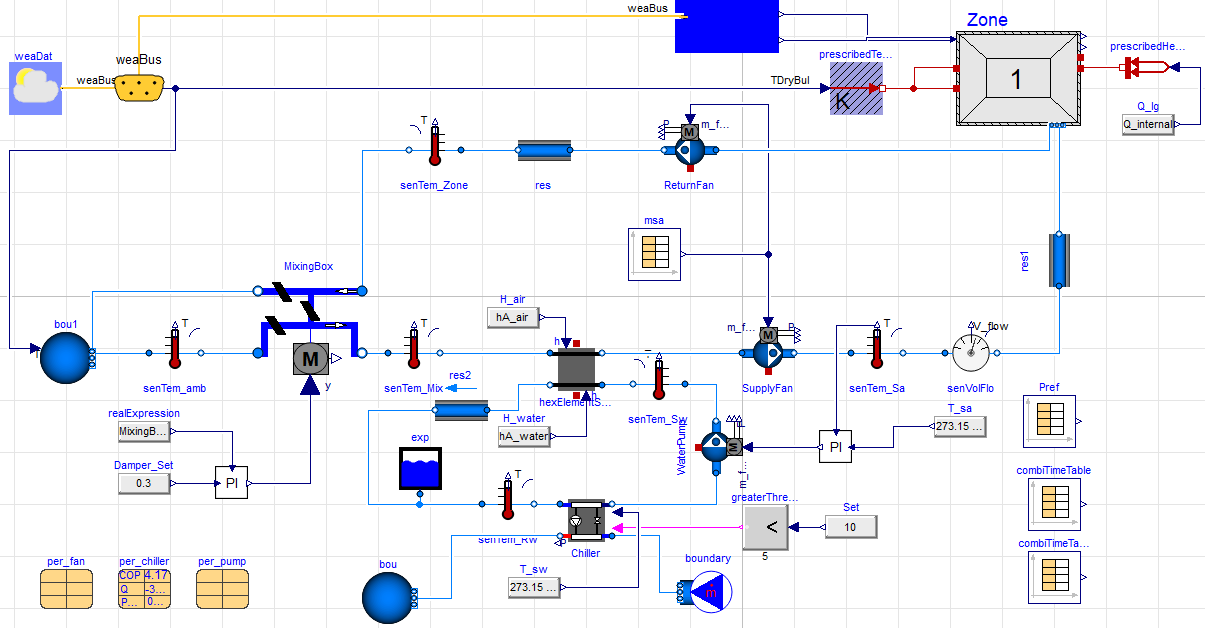}}
\end{center}
\caption{One air-handling-unit (AHU) thermal zone Dymola simulation model}
\label{dymola}
\end{figure}

The physical model-based thermal dynamic simulations of the building-HVAC system are performed by Dymola in Fig. \ref{dymola}. It serves as the ground truth model for evaluating realistic system outcomes, such as power consumption and zone temperature. 

\subsection{HVAC Power Tracking Baseline Scenarios}\label{baseline}
\begin{figure}[h]
\begin{center}
{\includegraphics[width=0.5\textwidth]{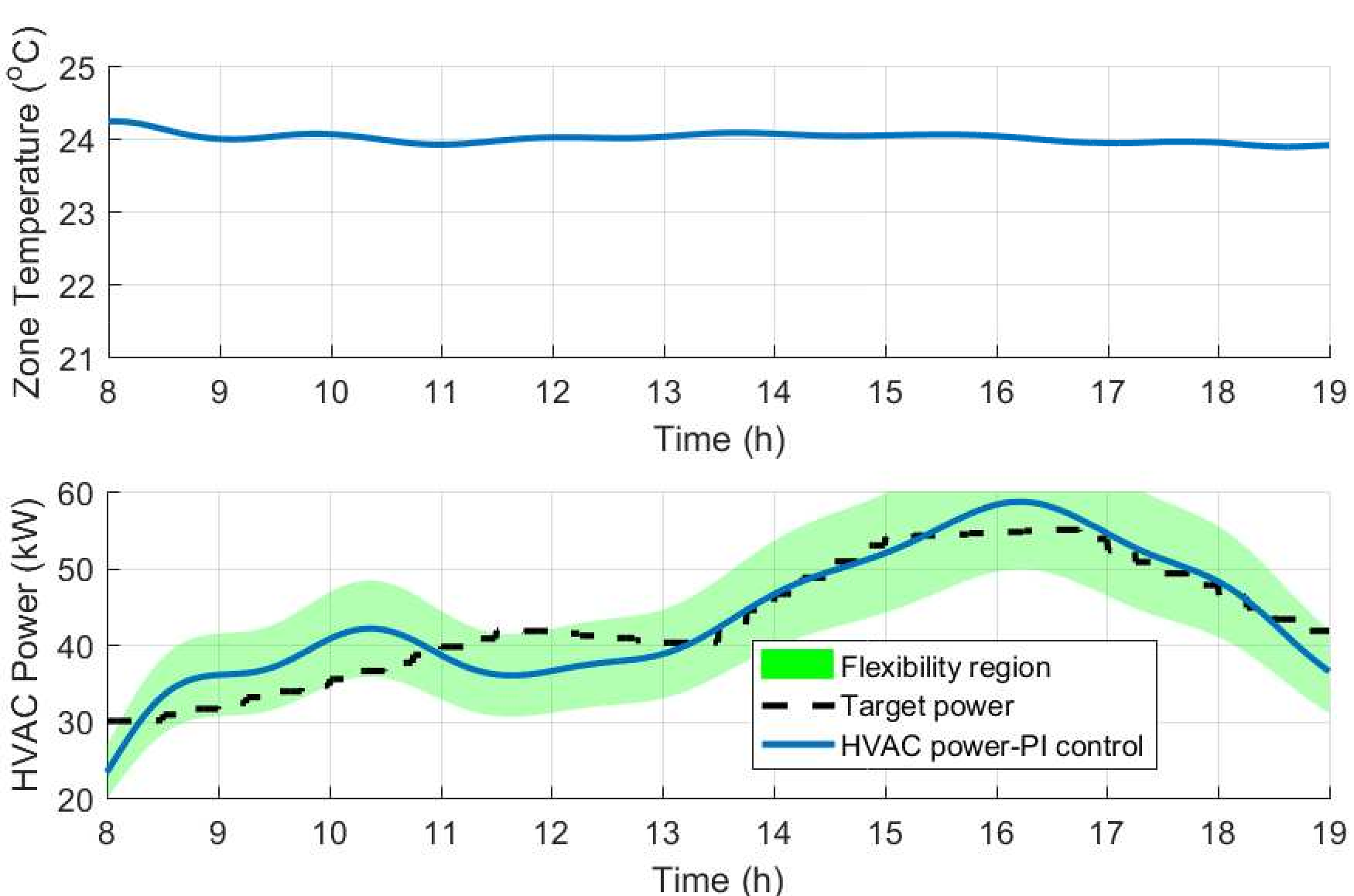}}
\end{center}
\caption{HVAC system outcomes under PI control}
\label{dailyPI}
\end{figure}

The traditional PI-controlled HVAC system performance is set as the baseline scenario. Its temperature and load profiles are provided in Fig. \ref{dailyPI}, where the zone temperature is closely maintained at the nominal setpoint of \SI{24}{\celsius}, and the power profile is fixed to the blue solid trajectory. The shaded green area denotes a $\pm 15\%$ flexibility region. The target power profile is the dashed trajectory that has a lower peak and higher valley load.  

\begin{figure}[h]
\begin{center}
{\includegraphics[width=0.5\textwidth]{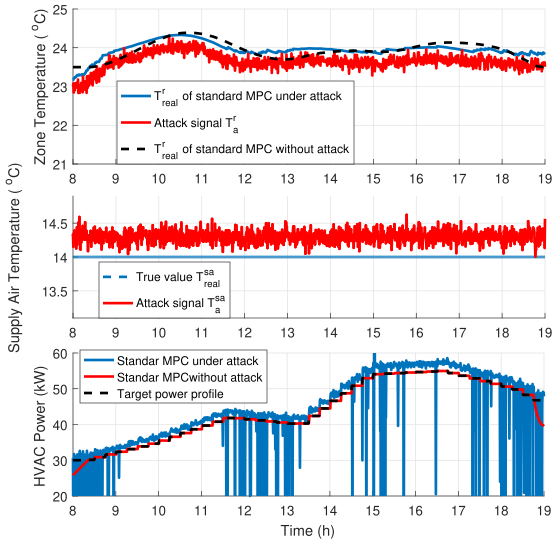}}
\end{center}
\caption{Power and temperature trajectories under standard MPC}
\label{dailyMPCbase2}
\end{figure}

The standard MPC is adopted to track the target power profile. The dashed power trajectory in Fig. \ref{dailyMPCbase2} denotes its numerically predicted profile, which can closely track the target one. The RMSE between these two profiles is 0.9159 kW, which reflects a high tracking accuracy. However, this good tracking performance can only be achieved under true sensor measurements. When the attack signals $T^r_a$ and $T^{sa}_a$ are fed into the standard MPC as sensor values, the resulting power profile of the standard MPC becomes the blue trajectory, which is highly oscillatory. The power tracking RMSE of this profile rises to 9.7563 kW. It is worth noting that, the most effective attack is to compromise the power tracking performance by creating large power dips occasionally while maintaining the power slightly higher than the target value for the most of the time.

\begin{figure}[h]
\begin{center}
{\includegraphics[width=0.5\textwidth]{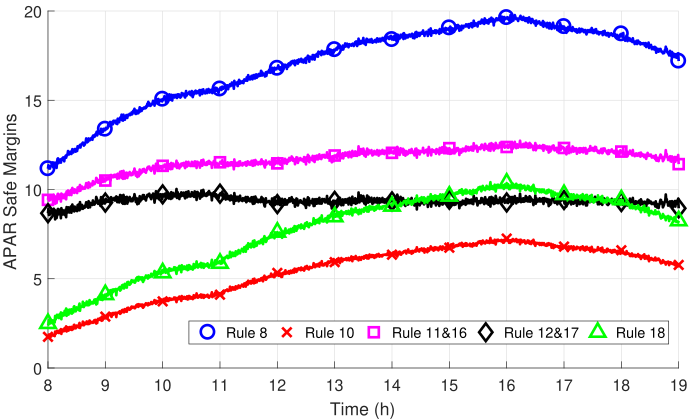}}
\end{center}
\caption{Fault detection safe margins under APAR}
\label{apar}
\end{figure}

While the power tracking performance is largely compromised, it will not be detected by the APAR method. Fig. \ref{apar} shows the APAR safe margins under the rules relevant to the two mechanical cooling modes, "Mode 3: Mechanical cooling with 100\% outdoor air" and "Mode 4: Mechanical cooling with minimum outdoor air". The APAR is triggered when the safe margin of any rule is below zero. However, none of the rules is triggered, indicating the attack can remain stealthy within the full operating range of the mixing box damper. 

\subsection{Performance of Resilient Control}
\begin{figure}[h]
\begin{center}
{\includegraphics[width=0.5\textwidth]{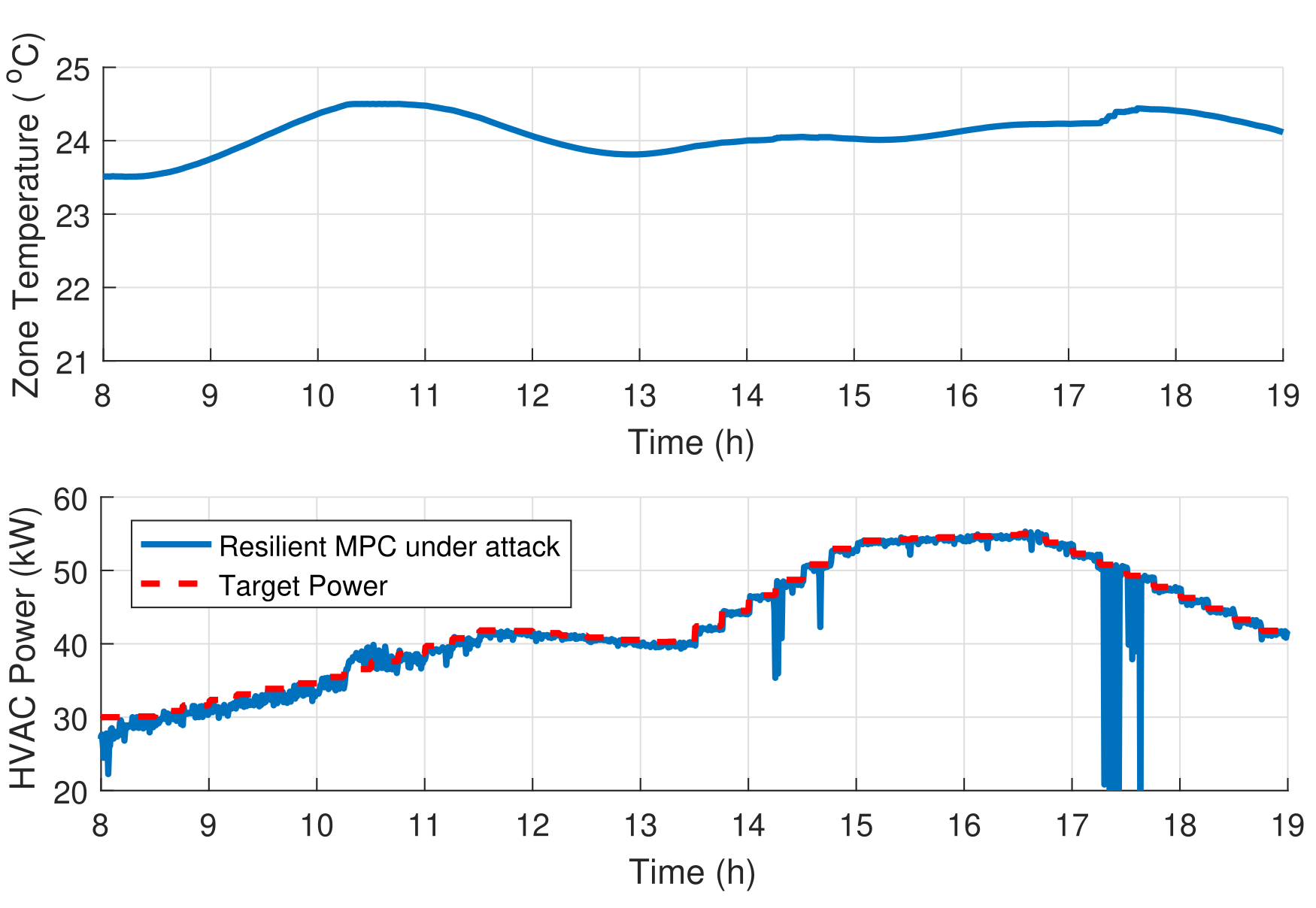}}
\end{center}
\caption{Performance of stealthy sensor attack under the daily test case}
\label{daily1}
\end{figure}

To defend against unknown stealthy sensor attacks, we substitute the standard MPC with the proposed resilient MPC. Fig. \ref{daily1} shows its performance under the same stealthy attack as in subsection \ref{baseline}. The power oscillations and deviations are noticeably alleviated, yielding a power tracking RMSE of 2.6426 kW, which is over 70\% lower than the standard MPC. The significantly reduced power tracking error indicates the high effectiveness of the proposed resilient control algorithm. Furthermore, the zone temperature is well maintained within the comfort zone around the setpoint of \SI{24}{\celsius}, i.e., the desired functionality of the HVAC system is not sacrificed.



In practice, both the stealthy sensor attack and resilient control are expected to be implemented in the real-time operation of the HVAC system. Thus, time efficiency is critical to the practicality of attack and defense. Table \ref{time} summarizes their average time consumption per step. The stealthy sensor attack problem is relaxed into a convex problem and can be solved extremely fast. The average time consumption per step is about 1 second, which includes 4 to 5 iterations. Although the resilient control problem is non-convex, the problem is most likely to be solved within 10 seconds with a good initialization. Considering the 30 seconds time step interval, both algorithms are fast enough for real-time implementation.

\begin{table}[ht!]
\begin{center}
\caption{Time Consumption per Step}
\label{time}
\scalebox{1}{
\begin{tabular}{ c | c c }
\hline
 & Attack & Resilient MPC \\
\hline
Mean (sec) & 1.07 & 6.31\\
\hline
Std & 0.20 & 1.95\\
\hline
\end{tabular}}
\end{center}
\end{table}

\begin{figure}[h]
\begin{center}
{\includegraphics[width=0.5\textwidth]{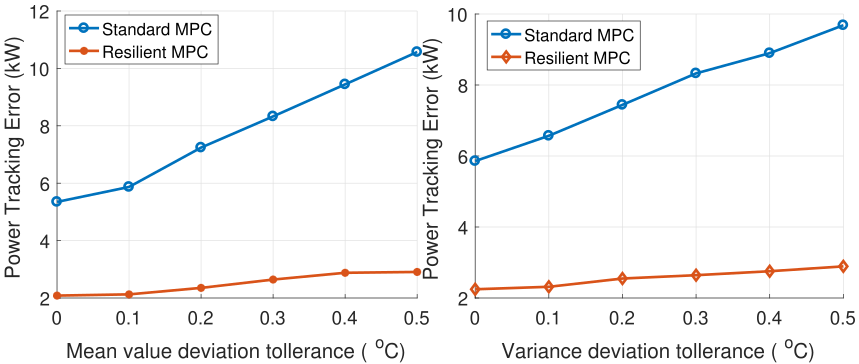}}
\end{center}
\caption{Robustness of resilient control}
\label{sensi}
\end{figure}

To further demonstrate the robustness of the proposed resilient control, a sensitivity study on the severity of the probabilistic attacks is performed. The severity is reflected by the mean and variance of the falsified attack signal. The higher values of these two parameters create a larger feasible region for the attacks. Correspondingly, a larger ambiguity set is required by the resilient MPC. Fig. \ref{sensi} shows that the power tracking error increases faster under standard MPC than under resilient MPC, which yields at most 3 kW power tracking error. It suggests that not only does the proposed resilient control yields lower power tracking errors than the standard MPC, but it also is more robust against the uncertainties of probabilistic attack parameters, which makes it more useful in practice. 

\subsection{Cross-validation using Dymola}

To cross-validate the above numerical simulation results, a digital twin model of the test building is developed in Dymola. The thermal and electric dynamic models in Dymola are all detailed first-principle models that can be regarded as very close to reality. Thus, it can serve as the ground truth model to validate the performance of the algorithms in reality. The environmental inputs are consistent with the numerical simulations. 

\begin{figure}[h]
\begin{center}
{\includegraphics[width=0.5\textwidth]{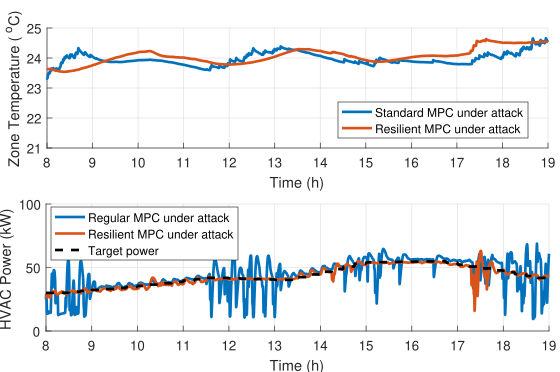}}
\end{center}
\caption{Performances of stealthy sensor attack and resilient control in Dymola}
\label{dymola2}
\end{figure}

Fig. \ref{dymola2} shows the daytime power and room temperature profiles obtained from Dymola simulation. With more realistic damping characteristics in the physical models of the chiller and supply air fan, the power drops in the power profile of both control methods are less extreme under attacks. But the standard MPC power trajectory is still highly oscillatory. The resilient MPC power trajectory noticeably tracks the target power better, which is consistent with the numerical simulation results. The RMSEs of the two control methods in this Dymola test case are 9.1765 kW and 2.7062 kW respectively, which suggests that the proposed resilient MPC yields a 70.51\% power tracking reduction. The room temperature profiles are similar and well maintained near the setpoint of \SI{24}{\celsius}. The Dymola simulation results in both power tracking and temperature maintaining are consistent with the numerical simulation results, while the minor deviations come from the modeling error between the higher-order physical models in Dymola and the simplified models adopted in Matlab scripts. Thus, the proposed resilient MPC can be expected to effectively diminish HVAC system power tracking error under sensor attacks in practice. 

\section{Conclusions}\label{conclusions}
In this paper, the problem of HVAC power tracking for the grid services by grid-interactive buildings is investigated under cyber attack scenarios. The formulations of sensor attacks are discussed, especially the stealthy sensor attack that can bypass the abnormal detection of APAR. A two-level resilient control is proposed as the defense strategy against stealthy sensor attacks and its tractable formulation is derived. 
Case studies suggest that the proposed resilient control significantly reduces the power tracking error under stealthy sensor attacks by over 70\% with sufficient time efficiency. Moreover, the power tracking robustness is implied by the insensitivity of power tracking errors to the intensity of sensor attacks. The results of the numerical model are consistent with that of the digital twin model in Dymola, which serves as the ground truth model, showing a high fidelity of the numerical models adopted in the resilient control formulation. The future work is to investigate the grid impact of coordinated cyber attacks on grid-interactive buildings and the corresponding defense strategies.

\begin{appendices}
\section{Building and HVAC Physical Models}\label{app1}
\subsubsection{HVAC Power Model}
The constant coefficient-of-performance (COP) model formulated in \eqref{coilpower} is a commonly adopted linear approximation of HVAC power \cite{raman2020model}. 
$Q_{coil}$ denotes the cooling coil thermal load and $COP$ is a constant that denotes the energy efficiency of a chiller.
\begin{equation}
P_c = \frac{Q_{coil}}{COP}
\label{coilpower}
\end{equation}

The cooling coil thermal load can be calculated from the heat gap between mixed air and supply air using \eqref{coilpower2}, where $c_{air}$ is a constant that represents the specific heat capacitance of air. $T^{mix}$ and $T^{sa}$ are the temperature of mixed air and supply air \cite{wang2022control}. 
\begin{equation}
Q_{coil} = c_{air}\dot{m}^{sa}\left(T^{mix} - T^{sa}\right)
\label{coilpower2}
\end{equation}

Substituting the mix-box model \eqref{mixbox} into \eqref{coilpower2}, the formulation of cooling coil thermal load becomes \eqref{coilpower3}, where $\beta$ is the damper position and $T^o$ is the ambient temperature. 
\begin{equation}
T^{mix} = \beta T^o + \left(1-\beta\right)T^r
\label{mixbox}
\end{equation}

\begin{equation}
Q_{coil} = \left(\beta T^o-T^{sa}\right)c_{air}\dot{m}^{sa} + \left(1-\beta\right)c_{air}\dot{m}^{sa}T^r
\label{coilpower3}
\end{equation}

Consequently, the original chiller power model \eqref{coilpower} now becomes \eqref{pmodelfull}, which can be represented by the compact form \eqref{coilpower4} with the aggregated coefficients $b_1, b_2$ and $b_3$. The aggregated coefficients are summarized in \eqref{coilpower5}. $b_1$ and $b_2$ are time varying because $T^o$ and $\beta$ are time varying inputs.
\begin{equation}
P_t = \beta T^o_t\frac{c_{air}}{COP}\dot{m}^{sa}_t + \left(1-\beta\right)\frac{c_{air}}{COP}\dot{m}^{sa}_tT^r_t - \frac{c_{air}}{COP}\dot{m}^{sa}_tT^{sa}_t
\label{pmodelfull}
\end{equation}

\begin{equation}
P_t = b_1\dot{m}^{sa}_t + b_2\dot{m}^{sa}_tT^r_t + b_3\dot{m}^{sa}_tT^{sa}_t
\label{coilpower4}
\end{equation}

\begin{equation}
\left\{
\begin{aligned}
b_1 &= \beta T^o_t\frac{c_{air}}{COP}\\
b_2 &= \left(1-\beta\right)\frac{c_{air}}{COP}\\
b_3 &= -\frac{c_{air}}{COP}
\end{aligned}
\right.
\label{coilpower5}
\end{equation}

\subsubsection{Thermal Zone Temperature Model}
\begin{figure}[h]
\begin{center}
{\includegraphics[width=0.3\textwidth]{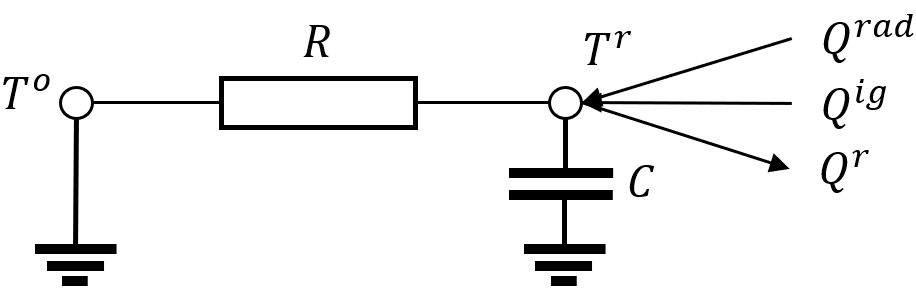}}
\end{center}
\caption{Thermal zone RC equivalent model}
\label{rcmodel}
\end{figure}

The thermal zone temperature model is derived from the RC equivalent model in Fig. \ref{rcmodel}, where $R$ denotes the thermal resistance of walls and $C$ denotes the thermal capacitance of room air mass. The thermal injection of solar radiation $Q^{rad}$ is mainly gained through windows, and the internal heat gain $Q^{ig}$ is obtained from occupancy, lighting, and miscellaneous electric loads. $Q^{r}$ denotes the heat removed from the room by the HVAC system, i.e., the thermal gap between the air entering and leaving the room as formulated in \eqref{Qr}.

\begin{equation}
Q^r = c_{air}\dot{m}^{sa}\left(T^r-T^{sa}\right) 
\label{Qr}
\end{equation}

The law of thermal conservation of a thermal zone is modeled by the KCL at the thermal zone node in \eqref{tmodeldiff}, which can be discretized into \eqref{tmodeldiff2}. $\Delta t$ denotes the step size.
\begin{subequations}
\begin{align}
C\frac{dT^r}{dt} &= \frac{T^o-T^r}{R} - Q^{r} +Q^{ig} + Q^{rad} \label{tmodeldiff}\\
C\frac{T^r_{t+1}-T^r_t}{\Delta t} &= \frac{T^o_t-T^r_t}{R} - Q^{r}_t +Q^{ig}_t + Q^{rad}_t \label{tmodeldiff2}
\end{align}
\end{subequations}

Substituting \eqref{Qr} into \eqref{tmodeldiff2}, the zone temperature dynamic model is obtained in \eqref{tmodelfull}, which can be represented by the compact form \eqref{tmodel2} with the aggregated coefficients $c_0\sim c_4$ as summarized in \eqref{tmodel3}. $c_0$ is time varying because $Q^{ig},Q^{rad}$, and $T^o$ are time varying inputs.
\begin{equation}
T^r_{t+1} = \left(1-\frac{\Delta t}{RC}\right)T^r_t -\frac{\Delta t c_{air}}{C} \dot{m}^{sa}_{t}T^r_t  + \frac{\Delta t c_{air}}{C}\dot{m}^{sa}_{t}T^{sa}_t + c_0
\label{tmodelfull}
\end{equation}

\begin{equation}
T^r_{t+1}= c_0 + c_1T^r_t + c_2 \dot{m}^{sa}_{t}T^r_t + c_3\dot{m}^{sa}_{t}T^{sa}_t
\label{tmodel2}
\end{equation}

\begin{equation}
\left\{
\begin{aligned}
c_0 &= \frac{\Delta t}{RC}T^o_t + \frac{\Delta t}{C}\left(Q^{ig}_t+Q^{rad}_t\right)\\
c_1 &= 1-\frac{\Delta t}{RC}\\
c_2 &= -\frac{\Delta t c_{air}}{C}\\
c_3 &= \frac{\Delta t c_{air}}{C}
\end{aligned}
\right.
\label{tmodel3}
\end{equation}

\section{Vectors and Matrices in Compact Formulations}\label{app2}
\begin{equation}
\allowdisplaybreaks
c_t = \left[\begin{aligned}
2a_4a_5\dot{m}^{sa^2}_t&-2a_5P^{ref}_t\dot{m}^{sa}_t\\
&a_5^2\dot{m}^{sa^2}_t \\
2a_4a_6\dot{m}^{sa^2}_t&-2a_6P^{ref}_t\dot{m}^{sa}_t\\
&a_6^2\dot{m}^{sa^2}_t \\
&2a_5a_6\dot{m}^{sa^2}_t 
\end{aligned}\right]
\label{ct}
\end{equation}

\begin{figure*}
\begin{equation}
\allowdisplaybreaks
A_t = \left[\begin{aligned}
\left(a_1+a_2\dot{m}^{sa}_t\right)&\qquad 0&a_3\dot{m}^{sa}_t& \qquad 0&0\qquad\\
2a_0\left(a_1+a_2\dot{m}^{sa}_t\right)& \quad \left(a_1+a_2\dot{m}^{sa}_t\right)^2&2a_0a_3\dot{m}^{sa}_t&\quad a_3^2\dot{m}^{sa^2}_t&2\left(a_1+a_2\dot{m}^{sa}_t\right)a_3\dot{m}^{sa}_t\\
0&\qquad0&1&\qquad0&0\\
0&\qquad0&0&\qquad1&0\\
0&\qquad0&a_0& \qquad a_3\dot{m}^{sa}_t&a_1+a_2\dot{m}^{sa}_t\\
\end{aligned}\right]
\label{A}
\end{equation}
\end{figure*}

\begin{equation}
\allowdisplaybreaks
B = \left[\begin{aligned}
&a_0\\
&a_0^2\\
&0\\
&0\\
&0\\
\end{aligned}\right], \quad  
D_t = \left[\begin{aligned}
&T^r_{ub}\\
&T^r_{t}+\varepsilon\\
&T^{sa}_{t}+\gamma\\
&-T^r_{lb}\\
&\varepsilon - T^r_{t}\\
&\gamma - T^{sa}_{t}
\end{aligned}\right]
\label{B}
\end{equation}

\begin{equation}
\allowdisplaybreaks
C = \left[\begin{aligned}
1 &\qquad0 &0 &\qquad0 &0\\
1 &\qquad0 &0 &\qquad0 &0\\
0 &\qquad0 &1 &\qquad0 &0\\
-1 &\qquad0 &0 &\qquad0 &0\\
-1 &\qquad0 &0 &\qquad0 &0\\
0 &\qquad0 &-1 &\qquad0 &0\\
0 &\qquad1 &0 &\qquad0 &0\\
0 &\qquad0 &0 &\qquad1 &0\\
\end{aligned}\right]
\label{C}
\end{equation}

\begin{equation}
c' = \left[c_1^T \quad c_2^T\quad ... \quad c_T^T \quad 0\quad 0\quad ...\quad 0\right]^T
\label{cprime}
\end{equation}

\begin{equation}
\resizebox{\hsize}{!}{
\allowdisplaybreaks
$A^\prime = \left[\begin{aligned}
A_1 &\quad -I & \textbf0 &\qquad\textbf0 &\quad\textbf0&\quad\cdots&\quad\cdots&\quad\textbf0\\
\textbf0&\quad\ddots& \ddots&\qquad\textbf0 &\quad\vdots&\quad\ddots&\quad\ddots&\quad\vdots\\
\textbf0 &\qquad \textbf0 & A_{T-1} &\quad -I &\quad\textbf0&\quad\cdots&\quad\cdots&\quad\textbf0\\
C &\qquad \textbf0 & \cdots &\qquad \textbf0 &\quad I&\qquad\textbf0&\quad\cdots&\quad\textbf0\\
\textbf0 &\qquad C_{\left(1,4,:\right)} & \ddots &\qquad \vdots &\quad\textbf0&\quad\ddots&\quad\ddots&\quad\vdots\\
\vdots &\qquad \ddots & \ddots &\qquad \textbf0 &\quad\vdots&\quad\ddots&\quad\ddots&\quad\textbf0\\
\textbf0 &\qquad \cdots & \textbf0 &\qquad C_{\left(1,4,:\right)} &\quad\textbf0&\quad\cdots&\quad\textbf0&\quad I\\
\end{aligned}\right]$
}
\label{Aprime}
\end{equation}

\begin{equation}
B' = \left[B^T \quad ... \quad B^T \quad D^T\quad D_{t\left(1,4\right)}^T\quad ...\quad D_{t\left(1,4\right)}^T\right]^T
\label{Bprime}
\end{equation}


\end{appendices}

\bibliographystyle{IEEEtran}
\bibliography{HVAC.bib}

\end{document}